# The Starburst-Driven Molecular Wind in NGC 253 and the Suppression of Star Formation

Alberto D. Bolatto[1], Steven R. Warren[1], Adam K. Leroy[2], Fabian Walter[3], Sylvain Veilleux[1], Eve C. Ostriker[4], Jürgen Ott[5], Martin Zwaan[6], David B. Fisher[1], Axel Weiss[7], Erik Rosolowsky[8,9], and Jacqueline Hodge[3]

[1] Department of Astronomy, Laboratory for Millimeter-wave Astronomy, and Joint Space Institute, University of Maryland, College Park, MD 20742, USA

[2] National Radio Astronomy Observatory, Charlottesville, VA 22903, USA

[3] Max-Planck Institut für Astronomie, Königstuhl 17, D-69117, Heidelberg, Germany

[4] Department of Astrophysical Sciences, Princeton University, Princeton, NJ 08544, USA

[5] National Radio Astronomy Observatory, P.O. Box O, 1003 Lopezville Road, Socorro, NM 87801, USA

[6] European Southern Observatory, Karl-Schwarzschild-Str. 2, Garching 85748, Germany

[7] Max-Planck-Institut für Radioastronomie, Auf dem Hügel 69, 53121 Bonn, Germany

[8] Department of Physics and Astronomy, University of British Columbia, Okanagan, Kelowna BC V1V 1V7, Canada

[9] Currently at Department of Physics, University of Alberta, Edmonton, Alberta T6G 2E1, Canada

**The under-abundance of very massive galaxies[1,2] in the universe is frequently attributed to the effect of galactic winds[3,4,5,6]. Although ionized galactic winds are readily observable most of the expelled mass is likely in cooler atomic[7,8] and molecular phases[9,10,11]. Expanding molecular shells observed in starburst systems such as NGC 253[12] and M 82[13,14] may facilitate the entrainment of molecular gas in the wind. While shell properties are well constrained[12], determining the amount of outflowing gas emerging from such shells and the connection between this gas and the ionized wind requires spatial resolution <100 pc coupled with sensitivity to a wide range of spatial scales, hitherto not available. Here we report observations of NGC 253, a nearby[15] starburst galaxy (D~3.4 Mpc) known to possess a wind[16,17,18,19,20], which trace the cool molecular wind at 50 pc resolution. At this resolution the extraplanar**

molecular gas closely tracks the Hα filaments[18], and it appears connected to molecular expanding shells[12] located in the starburst region. These observations allow us to directly measure the molecular outflow rate to be $dM/dt > 3$ M$_\odot$ yr$^{-1}$ and likely ~9 M$_\odot$ yr$^{-1}$. This implies a ratio of mass-outflow rate to star formation rate of at least η~1-3, establishing the importance of the starburst-driven wind in limiting the star formation activity and the final stellar content.

The ionized wind from NGC 253 emerges from its central ~200 pc (Figure 1) and has a central axis approximately in the plane of the sky ($i\approx78°$) filling a cone of 60° opening angle, with inclination-corrected outflow speeds[18] of a few hundreds km s$^{-1}$. Likely because of obscuration this central wind region is highly asymmetric, with Hα emission predominantly on the approaching (front) side of the outflow. Outflow activity extends to ~10 kpc, with lobes of Hα, X-ray, and UV emission evident above and below the plane of the galaxy[17]. Although it is unclear whether NGC 253 hosts an accreting black hole[22,23], the corresponding low-luminosity active galactic nucleus is not energetically dominant. The NGC 253 outflow has the characteristic low velocity of starburst-driven winds[21], and is almost certainly driven purely by star formation. HST imaging of the central region reveals absorption lanes due to dust entrained in the outflow, suggesting the wind may also carry significant amounts of molecular gas. Imaging of the cold molecular wind itself, however, has not been performed.

We imaged the $^{12}$CO $J$=1-0 transition in the central arcminute of NGC 253 using the Atacama Large Millimeter Array (ALMA) located in northern Chile. These data were combined with a single-dish map obtained at Mopra. The details of the observation and processing of the data can be found in the Supplementary Information. The resulting data cube has a sensitivity of 6 mJy beam$^{-1}$ (~54 mK) in 2.5 km s$^{-1}$ and an angular resolution of $\theta$~3.2''. These data have order-of-magnitude better Rayleigh-Jeans temperature sensitivity than previous $^{12}$CO interferometric imaging[12,24] of NGC 253 at its published resolution. This is not the first molecular wind imaged[9,10,11], but these observations make possible the study of the outflow in unprecedented detail, allowing us to map the spatially resolved structure of the cold phase of the galactic wind using a tracer that can be directly related to its mass.

The imaging reveals the presence of previously unknown low-level emission features that are approximately perpendicular to the bright lane of CO emission in the central region of this barred galaxy (Figure 2). These molecular streamers have surface brightness in the 30-200 mJy beam$^{-1}$ range. Perhaps the most prominent feature is a ridge of CO emission at v~70-250 km s$^{-1}$ emerging on the SW edge of the optically bright nuclear region, coincident with a linear dark dust feature along the western edge of the Hα and X-ray outflow (Figures 2 and 3). CO emission also extends south from the eastern regions of the nucleus at $v$~40-140 km s$^{-1}$, which we associate with the SE component of the molecular wind. These features trace an arc almost perfectly aligned with the edges of the Hα outflow (Figure 3), making evident the close spatial correlation between the high emission-measure ionized filaments and the molecular gas. The receding side of the galactic wind, invisible in Hα and barely hinted at in X-ray emission due to absorption from the intervening disk, is apparent in our CO images at velocities $v\sim$240-400 km s$^{-1}$ as a two-pronged structure located NW of the nucleus. We see no clear evidence for a corresponding NE component.

Several of the extraplanar CO features can be traced back in position and velocity to molecular expanding shell structures in the starburst region of NGC 253, providing clues to the launching mechanisms of the molecular wind (Figure 2). Two of these structures (shells 1 and 3) were found by previous CO observations and dubbed "superbubbles"[12]. We find four expanding shells with radii $r_{sh}\sim$60-90 pc, expansion velocities $V_{sh}\sim$23-42 km s$^{-1}$, characteristic dynamical ages $t_{sh}\sim$1.4-4 Myr, and molecular masses $M_{sh}\sim$0.3-1×10$^7$ M$_\odot$ (see Supplementary Information). We measure momenta $p_{sh}\sim$8.5-40×10$^7$ M$_\odot$ km s$^{-1}$, and energies $E_{sh}\sim$2-20×10$^{52}$ erg. The large momentum involved in each shell suggests driving by the combined effects of multiple stellar winds and (at later stages) supernovae originating in young stellar clusters. These clusters are not directly observed except perhaps for shell 1 (see below). Assuming a steadily driven superbubble[25] with mechanical power $L\sim$10$^{39}$ erg s$^{-1}$ ($M_{cluster}$/10$^5$ M$_\odot$), from the shell momenta we infer stellar cluster masses $M_{cluster}\sim$6×10$^4$-4×10$^5$ M$_\odot$ if the typical cluster age is $t_{cluster}\sim$3 Myr (i.e., before supernovae contribute), or lower by a factor $\sim$($t_{cluster}$/3 Myr)$^{-1.75}$ for older clusters. Shell 1 corresponds to superbubble SB2[12], and contains a number of sources suggestive of compact clusters in the *Hubble Space Telescope* WFC3 Paschen β archival image (PI M. Westmoquette). Shell 3 corresponds approximately to superbubble SB1[12], and exhibits [FeII] 1.644 μm emission[26] coincident with a radio continuum compact source[27], presumably a supernova remnant. Shells 2 and 4 are located in regions of very high extinction, even in the infrared, that contain no sources in the HST images. It is unclear whether shells 3 and 4 are related to features in the wind. Simulations of stellar feedback and galaxy winds find that supernovae, stellar winds, and radiation pressure reinforce each other synergistically[6]. We appear to be seeing these mechanisms in action in NGC 253, where the modest expansion velocities of the observed shells likely impart the initial momentum to the molecular gas, which is then advected into the hot outflowing material.

We use the observed CO luminosities and velocities to estimate the mass, mass loss rate, and energetics of the molecular wind. The CO luminosity of the four different components identified in the molecular wind (SW, SE, NW1, and NW2) adds up to $L_{CO}\approx$ 2.0×10$^7$ K km s$^{-1}$ pc$^2$, approximately equally split between the S and N portions of the outflow. To compute molecular masses in the wind, we adopt an optically thin conversion factor $\alpha_{CO}\approx$0.34 M$_\odot$ (K km s$^{-1}$ pc$^2$)$^{-1}$, about an order of magnitude lower than the value characteristic of the Milky Way disk (Supplementary Information). This results in a total outflowing molecular mass $M_{mol}\sim$6.6×10$^6$ M$_\odot$, including the correction for He abundance. The observed projected velocities for the CO wind components are $v_w\sim$30-60 km s$^{-1}$ relative to the nearby emission in the nuclear region, with a large de-projection correction due to the low inferred inclination[18] of the outflow (Supplementary Information). These velocities are lower than those of the ionized component[18], suggesting that advection plays an important role. The projected filament lengths are $r_w \sim$ 120-320 pc, and the implied dynamical filament ages $t_{dyn} \sim$ 0.3-1 Myr.

The resulting total molecular mass outflow rate is $\dot{M}_w \sim$ 9 $M_\odot$ yr$^{-1}$ with considerable uncertainty due to $\alpha_{CO}$ and the geometrical corrections. Note, however, that most likely corrections would yield an increase in the outflow rate over this value as our assumptions lead to a conservative estimate (Supplementary Information). Adopting the $\alpha_{CO}$ used for the central regions of NGC 253 would increase the outflow rate to $\dot{M}_w \sim$ 30 $M_\odot$ yr$^{-1}$, which we consider a likely upper limit. The star formation rate in the starburst of NGC 253 is $SFR \sim$ 2.8 $M_\odot$ yr$^{-1}$, determined from radio continuum and far-infrared measurements[28]. Obtaining an outflow rate similar to the SFR would require decreasing the CO excitation to $T_{ex} \sim$ 10 K, increasing the CO abundance by a factor of 3, increasing the inclination of the outflow to $i \sim$ 45°, or a combination of the above. These are possible but extreme corrections for a variety of reasons (Supplementary Information). We conclude that $\dot{M}_w \sim$ 3 $M_\odot$ yr$^{-1}$ is the lower bound of the possible molecular outflow rate values, implying as a robust result that the wind mass loss is at least equal to the SFR and likely a few times higher.

Consequently, the mass loading of the wind is certainly $\eta \geq 1$ from our lower limit, and likely $\eta \sim 3$. Recent estimates based on modeling of unresolved OH absorption spectroscopy[19] suggest $\dot{M}_w \sim$ 1.6 $M_\odot$ yr$^{-1}$ and possibly as large as 6.4 $M_\odot$ yr$^{-1}$, in approximate agreement with our more direct measurements. Over 90% of the CO luminosity of NGC 253 is found[29] within $R < 1'$ (~1 kpc), and even after taking into account variations in $\alpha_{CO}$ between the disk and the starburst the majority of the molecular gas is located in this region. Consequently, the central regions of NGC253 will run out of $H_2$ in ~60-120 Myr.

In general, it is not clear what fraction of the outflowing gas actually escapes galaxies, particularly for low velocity starburst-driven winds like that in NGC 253. Most of the baryons ejected by winds may just linger in the enriched halos of star forming galaxies[30], to later rain back to their disks providing fuel for new episodes of star formation. This recycling of baryons constitutes a third mode of galaxy accretion[5], which may be particularly important in shaping the galaxy mass function at intermediate and large galaxy masses[4]. Our work shows that the mass loading of the starburst-driven wind is substantial, supporting the importance of recycling. With the advent of ALMA the capabilities for sensitive imaging of the molecular component of galactic winds are dramatically improving, opening a new window onto the life cycle of baryons in galaxies.

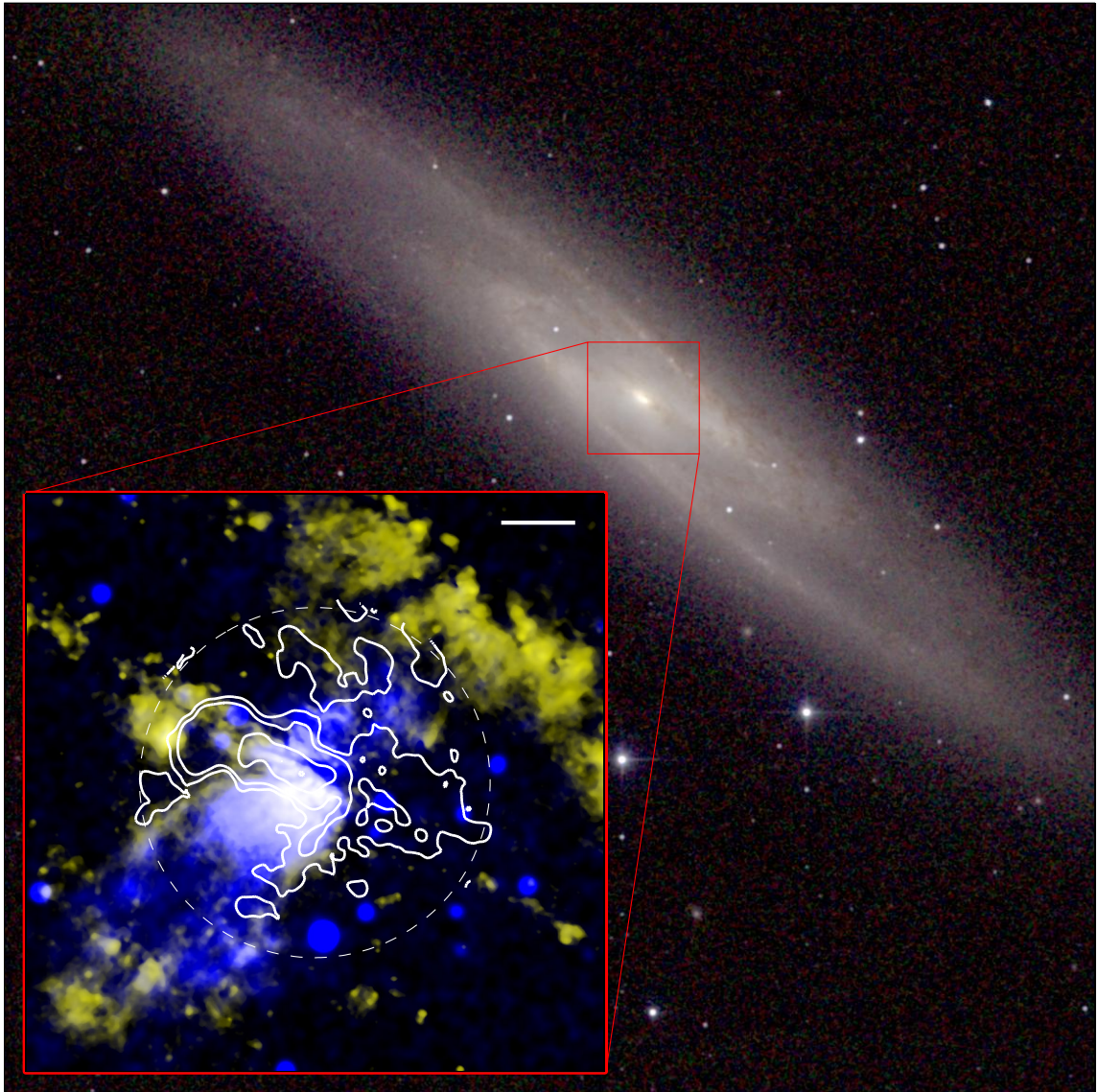

**Figure 1**: The warm and hot phases of the galactic wind in NGC 253. The background image shows the stellar disk of NGC 253 in the 2MASS JHK composite. The central 2 kpc are shown in the inset, in soft X-ray (blue) and Hα (yellow; obtained at CTIO and kindly provided by M. Lehnert) emission. The dashed white circle shows the field mapped with ALMA (centered on the reference position $\alpha_{2000}$=00$^h$47$^m$33$^s$.15, $\delta_{2000}$=-25°17'17".5) with a scale bar of 250 pc in the top right corner. The white contours represent CO emission at $V_{rad}$∼150 to 190 km s$^{-1}$ and correspond to $S_{CO}$=30, 120, 750, and 2500 mJy beam$^{-1}$ ($T_{CO}$≈0.25, 1, 7, 22 K), where the extraplanar streamers S of the central bar illustrate the front (approaching) side of the cold molecular wind (see Fig. 2; the contours N of the bar correspond to material at normal rotation velocities in the galaxy disk). The Hα emission includes material in the wind as well as emission from normal star forming regions in the bar and arms of NGC 253. The X-ray image from Chandra[16] shows emission from plasma in the wind as well as point sources in NGC 253 and the background.

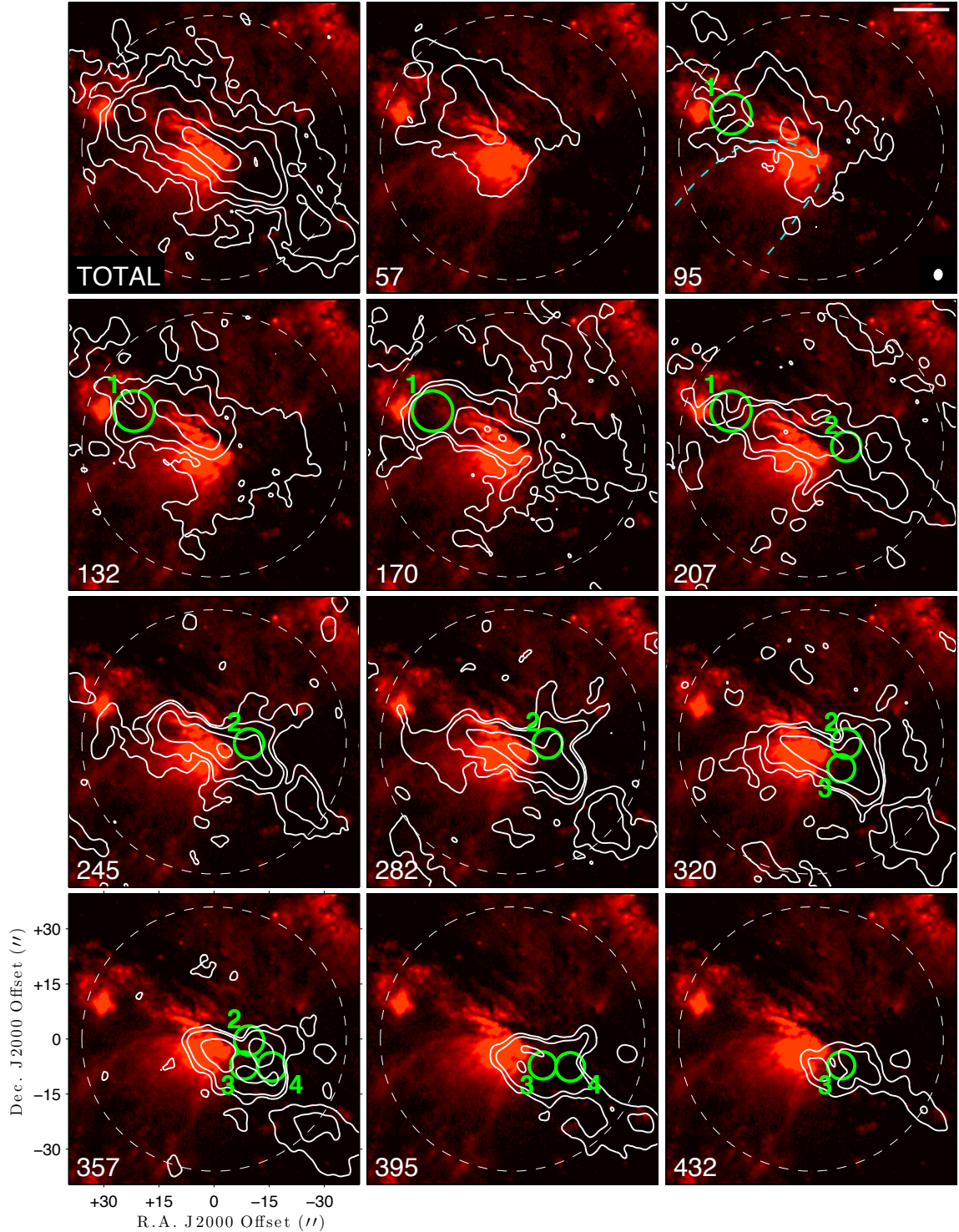

**Figure 2**: The molecular wind in NGC 253. Channel maps showing ALMA CO (1-0) observations against the Hα background image showing the outflow (North is up; East is left; reference coordinates are give in Fig. 1). The first panel shows the intensity integrated over all velocities ($S_{CO}\Delta v$=5, 10, 25, 100, 250 Jy km s$^{-1}$ beam$^{-1}$ contours). The other panels are channel maps spanning 37.5 km s$^{-1}$, with barycentric velocities indicated in the lower left corner of each panel ($S_{CO}$=30, 120,

750, and 2500 mJy beam$^{-1}$ contours). The panels are not corrected by field-of-view illumination, hence the signals are depressed near the edges of the mosaic. The white bar and ellipse in the top right panel represent respectively 250 pc, and the synthesized ALMA beam size (θ≈3.4"×3"). The dotted circle is the approximate half-power field of view of our 7-pointing mosaic. The green circles numbered 1 to 4 show the location and extent of the four expanding molecular shells identified in the CO cube. The dashed cyan line sketch shown in the channel at 96 km s$^{-1}$ illustrates the location and shape of the outer bright filaments in the Hα outflow[18]. The systemic heliocentric velocity of NGC 253 is 243 km s$^{-1}$, and the emission in the northern regions at 171-208 km s$^{-1}$ corresponds to material rotating with the disk.

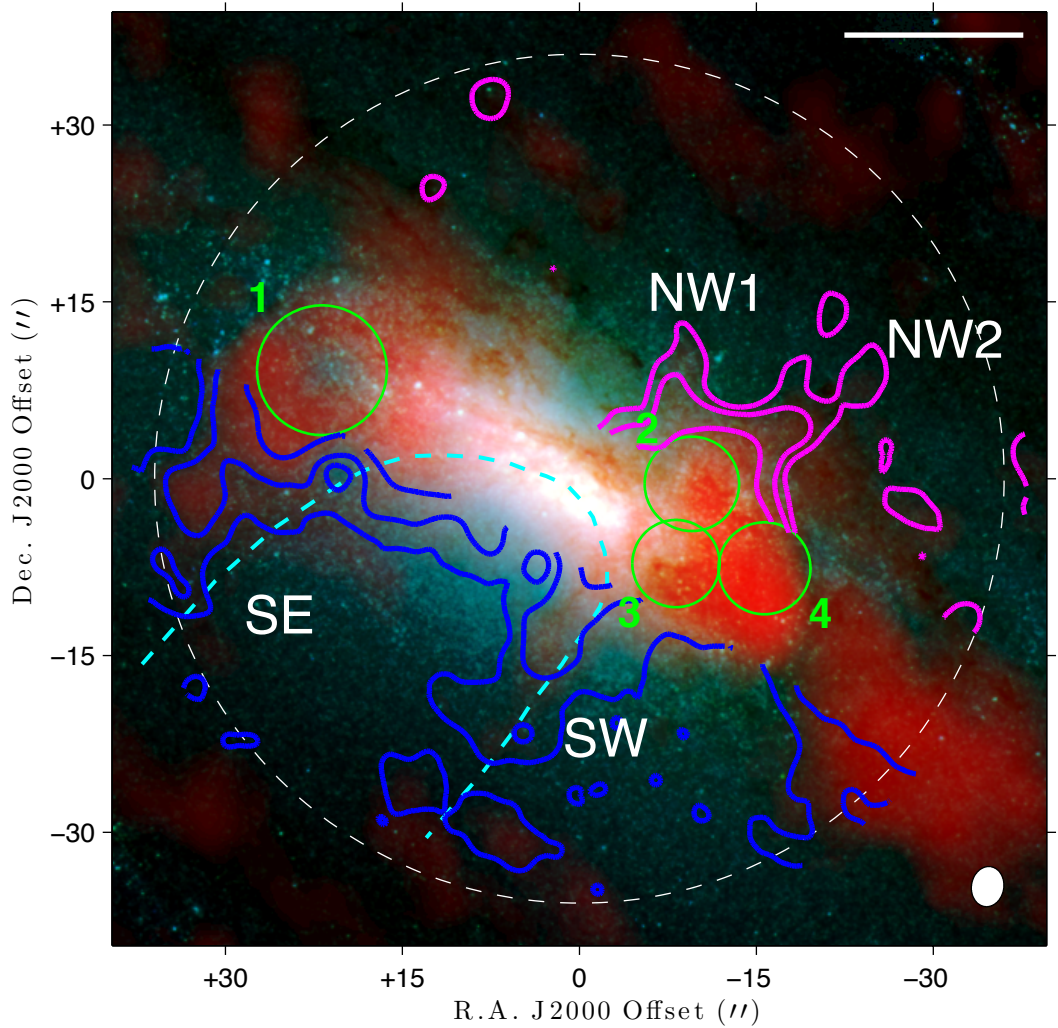

**Figure 3**: Integrated molecular wind emission in NGC 253. The background image is a color composite of HST J-band (blue), HST H-band (green), and ALMA CO (red) emission integrated over all velocities and corrected for the mosaic illumination (reference position is given in Fig. 1). The dotted circle, white bar, white ellipse, and dashed cyan line are as in Figure 1. The blue and magenta contours show the CO emission in the approaching (v≈73 to 273 km s$^{-1}$) and receding (v≈208 to 356 km s$^{-1}$) lobes of the outflow respectively. The region ±6" on each side of the galaxy plane

has been blanked for clarity. The contour levels correspond to $S_{CO}\Delta v$ = 5, 10, and 30 Jy km s$^{-1}$ beam$^{-1}$. The noise in the central regions of the mosaic is ~0.3 Jy km s$^{-1}$ beam$^{-1}$. The green circles illustrate the location and size of the expanding molecular shells.

**Acknowledgements** A.D.B. acknowledges partial support from a CAREER grant NSF-AST0955836, NSF-AST1139998, and from a Research Corporation for Science Advancement Cottrell Scholar award. S.V. acknowledges partial support through grant NSF-AST100958. E.C.O. is supported by the NSF through grant AST-0908185. ALMA is a partnership of ESO (representing its member states), NSF (USA) and NINS (Japan), together with NRC (Canada) and NSC and ASIAA (Taiwan), in cooperation with the Republic of Chile. The Joint ALMA Observatory is operated by ESO, AUI/NRAO and NAOJ. The National Radio Astronomy Observatory is a facility of the National Science Foundation operated under cooperative agreement by Associated Universities, Inc. We thank M. Lehnert for providing the Hα image, processed by himself and M. Dahlem.

**Author Contributions** A.D.B. and S.R.W. performed the detailed calculations used in the analysis. A.K.L., S.R.W., and A.D.B. reduced and analysed the ALMA data. D.B.F. reduced and analysed the HST archival data. A.D.B., F.W., A.K.L., and M.Z. wrote the ALMA proposal and designed the observations with input from coauthors. J.O. obtained and reduced the Mopra observations. A.D.B. wrote the manuscript with input from F.W., A.K.L., S.R.W., S. V., and E. C. O.. All authors were participants in the discussion of results, determination of the conclusions and revision of the manuscript.

**Author Information** Reprints and permissions information is available at www.nature.com/reprints. The authors declare no competing financial interests. Readers are welcome to comment on the online version of the paper. Correspondence and requests for materials should be addressed to A.D.B. (bolatto@astro.umd.edu). This paper makes use of the following ALMA data: ADS/JAO.ALMA#2011.0.00172.S.

# Supplementary Information

1. Observational Setup and Data Analysis

The ALMA receivers were tuned to the rest frequency of the ground rotational transition of carbon monoxide (CO), the most common tracer of cold molecular gas in galaxies. We used a 7-point mosaic pattern yielding a field of view with approximately flat sensitivity and a half-power radius of ~36". The interferometer configuration provided an angular resolution of 3.4"×3" oriented approximately N-S (using robust weighing with Briggs parameter 0.5), and the observations were designed to maximize their sensitivity to low surface brightness material. We attain an rms of ~6 mJy beam$^{-1}$ in 2.5 km s$^{-1}$ channels in thermal noise. Our dynamic range, however, is likely limiting the noise in the imaging in the presence of strong signal (our cycle 0 observations use ~16 antennas, offering limited sampling of the Fourier plane). We estimate the dynamic range to be ~250 with the brightest CO emission in the central region ~2-4 Jy beam$^{-1}$, suggesting an effective noise of 8-16 mJy beam$^{-1}$ per channel. We reduced the data using the CASA and Miriad software packages. After applying the phase calibrator solution we further iteratively self-calibrated the source on its continuum, which substantially reduced the phase noise. We combined the CLEANed ALMA observations with a single-dish map obtained at Mopra (a 22 m diameter dish located near Coonabarabran, Australia) using a feathering algorithm, to partially fill in the long spatial frequencies missing in the interferometric ALMA observations (shortest projected baseline ~20.2 m). The purpose of this correction is to mostly remove the "negative bowl" associated with missing spatial frequencies in the interferometer-only images, allowing us to recover extended flux. We then used the combined cube as the prior for a deconvolution of the ALMA observations using CLEAN.

## 2. Determination of the Molecular Gas Mass and Outflow Rate

Converting CO luminosity to a mass requires the adoption of a CO-to-H$_2$ conversion factor[31]. We use different conversion factors for the starburst disk of NGC 253, and for the molecular wind. The value of the conversion factor is reasonably well established for self-gravitating molecular clouds in galaxy disks, where $\alpha_{CO}\approx4.3$ M$_\odot$ (K km s$^{-1}$ pc$^2$)$^{-1}$, but it is not measured for a galactic wind. Modeling of the M 82 molecular outflow[32], however, suggests the CO emission originates in gas with low densities, low optical depth, and kinetic temperature of at least 30 K.

For the expanding molecular shells and other structures in the starbursting disk we adopt a CO-to-H$_2$ conversion factor $\alpha_{CO}\approx1$ M$_\odot$ (K km s$^{-1}$ pc$^2$)$^{-1}$. Note this factor includes the mass contribution of He mixed with the molecular gas. It is not uncommon for galaxy centers and starburst regions to emit more brightly in CO per unit H$_2$ mass than GMCs in the Milky Way, due to the combined effects of higher temperature and increased turbulent motions. Our adopted value is consistent with the modeling of the $^{13}$CO and $^{12}$CO emission[33] in NGC 253, which finds $X_{CO}\approx 0.5^{+0.4}_{-0.2}\times10^{20}$ cm$^{-2}$ (K km s$^{-1}$)$^{-1}$ in the innermost 2 kpc of this galaxy, or $\alpha_{CO}\approx1.1^{+0.8}_{-0.4}$ M$_\odot$ (K km s$^{-1}$ pc$^2$)$^{-1}$. This value also agrees well with a correction to the Galactic $X_{CO}$ values based on the measured stellar surface density[1]. We estimate the surface density in the central region using the HST archival H-band (F160W) image from WFC3. We derive it using the empirical relation[34] between NUV-J color index and M/L$_H$, which corrects for the effects of extinction. For a measured NUV-J≈7.31 in the innermost 40'', we expect log(M/L$_H$)≈-0.17±0.1. The measured H-band integrated flux over this area (m$_H$≈7.97) corrected by the observed inclination ($i$≈78°) then yields $\Sigma_*$~(3.5±1)×10$^3$ M$_\odot$ pc$^{-2}$. This suggests $\alpha_{CO}\approx0.7$ M$_\odot$ (K km s$^{-1}$ pc$^2$)$^{-1}$, consistent with the disk value adopted.

For the galactic wind we adopt an optically thin conversion factor[31], calculated using an excitation temperature $T_{ex} \approx 30$ K and an abundance $CO/H_2 \approx 10^{-4}$ implying $\alpha_{CO} \approx 0.34$ M$_\odot$ (K km s$^{-1}$ pc$^2$)$^{-1}$, about an order of magnitude lower than the value characteristic of the Milky Way disk. The optically thin conversion factor scales essentially linearly with $T_{ex}$, and is inversely proportional to $CO/H_2$. The assumption of optically thin emission leads to the lowest possible mass for the outflow, and is motivated by the likelihood of highly turbulent conditions in the emitting gas which would naturally lead to a decrease in the optical depth[31].

The choices of optically thin emission with this particular excitation temperature and CO abundance lead to a conservative estimate of the molecular mass in the outflow. In M 82 the kinetic temperature of the outflowing molecular gas is $T_{kin} \sim 30$-110 K with similar CO excitation temperatures[32], and we would expect molecular gas in the NGC253 wind to experience comparable conditions. Note also that the radiation temperature close to the starburst is likely to be high, and contribute significantly to raising $T_{ex}$ even if the gas densities are lower than the critical density of the CO ground transition ($n_{cr} \approx 2200$ cm$^{-3}$). Similarly, the $CO/H_2$ abundance ratio is likely to be lower in a galactic wind than in GMCs in the Milky Way, due to CO destruction caused by shocks or photodissociation. Both $T_{ex} > 30$ K and $CO/H_2 < 10^{-4}$ have the effect of increasing $\alpha_{CO}$ and the molecular mass of the wind.

To compute the mass outflow rate, $\dot{M}_w$, we use the equation
$$\dot{M}_w = \frac{2\alpha_{CO} L_{CO} v_w \tan i}{r_w},$$
where $L_{CO}$, $v_w$, $i$, and $r_w$ are respectively the CO luminosity, the projected velocity, the inclination with respect to the line-of-sight, and the projected length of the different outflow features ($r_w/2$ is the average distance traveled by the gas). Assuming the outflow is perpendicular to the disk ($i \approx 78°$) yields a large de-projection factor (tan $i \sim 4.7$). The 60° opening angle of the ionised outflow[18] suggests that it is possible for features to have inclinations as low as $i \sim 48°$ (tan $i \sim 1$). This extreme geometry, however, is highly unlikely. The location of the SE streamer, for example, following the edge of the ionised outflow cone, suggests its inclination is close to $i=78°$. The distribution of the de-projection factor for a cone centered on this inclination and with the observed opening angle has a median value of tan $i \sim 3$., which we adopt in our calculation (equivalent to $i \sim 72°$). Note that the distribution is skewed toward high values, so although values higher than 3 and lower than 3 are equally probable, the higher values can be much larger and would result in a proportionally larger mass outflow rate.